# 3D-XY critical fluctuations of the thermal expansivity in detwinned YBa$_2$Cu$_3$O$_{7-\delta}$ single crystals near optimal doping

Volker Pasler, Peter Schweiss and Christoph Meingast

*Forschungszentrum Karlsruhe, Institut für Nukleare Festkörperphysik,*

*P.O. Box 3640, 76021 Karlsruhe, Germany.*

Bernhard Obst and Helmut Wühl

*Forschungszentrum Karlsruhe, Institut für Technische Physik,*

*P.O. Box 3640, 76021 Karlsruhe, Germany.*

Alexandre I. Rykov and Setsuko Tajima

*Superconductivity Research Laboratory - ISTEC, 10 - 13 Shinonome 1- Chome, Koto-ku,*

*Tokyo 135, Japan.*

The strong coupling of superconductivity to the orthorhombic distortion in YBa$_2$Cu$_3$O$_{7-\delta}$ makes possible an analysis of the superconducting fluctuations without the necessity of subtracting any background. The present high-resolution capacitance dilatometry data unambiguously demonstrate the existence of critical, instead of Gaussian, fluctuations over a wide temperature region *(± 10 K)* around $T_c$. The values of the amplitude ratio $A^+ / A^- = 0.9 - 1.1$ and the leading scaling exponent $|\alpha| \leq 0.018$, determined via a least-squares fit of the data, are consistent with the 3D-XY universality class. Small deviations from pure 3D-XY behavior are discussed.

*PACS* numbers: 74.72.- h, 64.60.Fr, 65.70.+y

The superconducting transition in classical superconductors is described surprisingly well by mean-field theory. This is due to the large number of Cooper pairs in the coherence volume and is really an exceptional case for second-order transitions. In contrast, the small coherence lengths, high transition temperatures and quasi-2D nature of high- temperature superconductors (HTSC) greatly enlarge the temperature region in which fluctuations of the order parameter are important, and fluctuations in HTSCs have been observed in many types of experiments, e.g. specific heat [1-6], thermal expansion [7,8], resistivity [2,9], penetration depth [10,11] and magnetization [3,6,12] measurements. A large effort has gone into quantifying these effects, because a detailed analysis of the fluctuations can provide important information regarding the dimensionality and the order parameter of the superconducting state. The superconducting transition in the simplest case is expected to belong to the three-dimensional XY (3D-XY) universality class with a two-component order parameter, e.g. pure $d_{x^2-y^2}$ or $s$-wave. The most plausible order parameter that is most consistent with a large number of experiments on $YBa_2Cu_3O_{7-\delta}$, on the other hand, consists of a real mixture of $s^+$ and $d_{x^2-y^2}$ [13]. Here, we focus on the transition in the absence of an applied magnetic field, but note that many interesting phenomena, e.g. vortex fluctuations leading to first-order vortex melting, occur when a field is applied.

The fluctuation signal in most experiments is superimposed upon a large background, and it has proven quite difficult to determine whether the HTSCs are really in the critical regime or whether the fluctuations are better described by first-order corrections to mean-field theory, i.e. Gaussian fluctuations. For example, in specific heat measurements of $YBa_2Cu_3O_{7-\delta}$ the fluctuation component is at most 5 % of the large phonon background. By slightly adjusting the unknown phonon background, both critical and 3D-Gaussian type models can be fitted quite well [1,3,4-6]. The fitting residuals of the most recent specific heat measurements are somewhat smaller for the critical than for the Gaussian fits [3,4,6,14], although quite different backgrounds, with and without mean-field contributions, were used for the critical analyses. If a 10 K wide region around $T_c$ is excluded from the fits, critical and Gaussian fluctuations are no longer distinguishable [6,14]. Penetration depth measurements, which are ideally not plagued by any background, on $YBa_2Cu_3O_{7-\delta}$ appear to yield both critical [10] and mean-field type exponents [11, 15]. Thus, although there is increasing evidence for critical behavior in $YBa_2Cu_3O_{7-\delta}$ [4,6,9,10], the width of the critical region and the exact values of the critical exponents remain uncertain, and further experiments are desirable.

In this Letter, we present high-resolution thermal expansion data of untwinned YBa$_2$Cu$_3$O$_{7-\delta}$ single crystals near optimal doping. The thermal expansivity is thermodynamically closely related to the specific heat, and the same scaling laws apply close to $T_c$. This has been nicely demonstrated for the $\lambda$-transition in $^4$He [16]. For the present case, the advantage of the expansivity over the specific heat is that the size of the expansivity anomaly at $T_c$ relative to the background is about twenty times larger, making possible an analysis *without subtracting any background*. Our data unambiguously demonstrate the existence of critical, instead of Gaussian, fluctuations in a wide temperature region *(± 10 K)* around T$_c$. We show that the values of both the specific heat exponent $\alpha$ and the universal amplitude ratio $A^+ / A^-$ are consistent with the 3D-XY universality class. Small deviations from 3D-XY scaling are discussed.

Two YBa$_2$Cu$_3$O$_{7-\delta}$ single crystals (SRL-1 and SRL-2) grown by a pulling technique, detwinned at a uniaxial pressure of *10 MPa* and oxygenated at *490 °C* [17], were studied. Microscopic chemical characterization of the samples was carried out by energy dispersive x-ray spectroscopy (EDX). In this quantitative analysis, a high-purity YBa$_2$Cu$_3$O$_{7-\delta}$ standard, grown in a BaZrO$_3$ crucible [18], was used. Within the detection limit *(< 0.1 wt. %)*, no signs of impurities were found, demonstrating the exceptionally high purity of these crystals. An upper limit of 1% on the degree of twinning of crystal SRL-2 was obtained using neutron diffraction. The dimensions of the crystals ranged from 3-6 mm along the orthorhombic axes. The thermal expansion was measured with a high-resolution capacitance dilatometer upon heating at a rate of *5 mK/s*, and data points were taken every *20 mK* [7,19].

Fig. 1a shows the linear thermal expansivities $(\tilde{\alpha}_i(T) = 1 / L_i \cdot dL_i / dT; \ i = a,b,c)$ for the three orthorhombic axes of crystal SRL-1. (Unfortunately, the symbol $\alpha$ is usually used both for the thermal expansivity and the specific heat critical scaling exponent. We therefore use this somewhat unusual notation, $\tilde{\alpha}$, for the expansion coefficients in order to differentiate it from the critical scaling exponent $\alpha$.) Large anomalies of opposite sign are observed at $T_c$ in $\tilde{\alpha}_a$ and $\tilde{\alpha}_b$, whereas a small negative anomaly occurs in $\tilde{\alpha}_c$. These results are qualitatively similar to our previous data [7,19] and indicate that the crystal is near optimal doping [19]. The accuracy and resolution of the present measurements are, however, much better due to the larger crystal dimensions and improvements in the dilatometer. There is strong evidence that these anomalies are entirely due to superconductivity and not due to some additional structural effects. This is because the thermodynamic Ehrenfest relation, which relates the expansivity anomalies with the specific heat anomaly and the uniaxial pressure (stress) dependence of $T_c$, is satisfied very well [7,19,20].

For the purpose of looking at the fluctuations, it is very useful to consider the quantity $\tilde{\alpha}_{b-a} \equiv \tilde{\alpha}_b - \tilde{\alpha}_a$ (plotted in Fig. 1b) because the anomalies in $\tilde{\alpha}_a$ and $\tilde{\alpha}_b$ are added, whereas the background is reduced. This difference should also scale like the specific heat, since it is just a linear combination of two anomalies, each of which scales like $C_p$. The $\lambda$-character of the superconducting transition in this representation is clearly seen because the anomaly is of similar magnitude as the slowly varying background. As we demonstrate below, this allows an analysis without any background subtraction, as is possible for the $\lambda$-transition in $^4$He [15,21-23].

In the critical fluctuation region, the expansivity (specific heat) is expected to obey power laws of the form

$$\tilde{\alpha}_{b-a}(t>0) = \frac{A^+}{\alpha}|t|^{-\alpha} + C^+ \qquad\qquad -1-$$

$$\tilde{\alpha}_{b-a}(t<0) = \frac{A^-}{\alpha}|t|^{-\alpha} + C^- \qquad\qquad -2-$$

with the backgrounds $C^+ = C^-$ and $t \equiv (T / T_c - 1)$ [16,22,23]. The leading exponent $\alpha$ and the amplitude ratio $A^+ / A^-$ are universal quantities, i.e. they are the same for any system belonging to a given universality class, and both determine the shape of the transition close to $T_c$. In the 3D-XY universality class $\alpha$ is close to zero, and it is very difficult to distinguish a power law (Eq. 1 and 2) from a logarithmic divergence ($\alpha = 0$), as the specific heat experiments at the $\lambda$-point of $^4$He have demonstrated [16,21-23]. Since $\alpha$ and $A^+ / A^-$ are highly correlated, one needs very precise measurements over many decades of $t$ to accurately determine these quantities separately. For $^4$He, recent experiments performed in the space shuttle's gravitational free environment provide the best values ($\alpha = -0.01285 \pm 0.00038$, $A^+ / A^- = 1.054 \pm 0.001$), which are actually more precise than those determined by theory [23].

This same type of analysis in YBa$_2$Cu$_3$O$_{7-\delta}$ is unfortunately made quite difficult by the finite transition widths of even the best crystals. However, a closer inspection of Eq.s 1-2 in a $ln\ |t|$ representation suggests that the scaling at any point $|t_0|$ can also be characterized by the logarithmic derivatives $B^\pm = \dfrac{d}{d(ln|t|)}\left[\dfrac{A^\pm}{\alpha}|t|^{-\alpha}\right]_{t=t_0} = -A^\pm/t_0|^{-\alpha}$ and the difference $\Delta \tilde{\alpha}_{b-a} \equiv \tilde{\alpha}_{b-a}(t_o^-) - \tilde{\alpha}_{b-a}(t_o^+) = \left[ \dfrac{A^-}{\alpha} - \dfrac{A^+}{\alpha} \right] t_o \Gamma^{-\alpha}$ of Eq.s 1-2, respectively. In fact, the ratio of these quantities,

$$\Gamma \equiv \left( \frac{B^+ + B^-}{2} \right) / \Delta \tilde{\alpha}_{b-a} = \frac{\alpha}{2} \cdot \frac{\dfrac{A^+}{A^-} + 1}{\dfrac{A^+}{A^-} - 1}, \qquad \text{-3-}$$

depends only on the universal quantities $\alpha$ and $A^+ / A^-$. Thus, $\Gamma$ is also a universal scaling quantity for a given universality class. For ${}^4$He, the values of $\alpha$ and $A^+ / A^-$ from both specific heat [21-23] and thermal expansion studies [16] vary considerably, whereas the $\Gamma$ values are practically constant ($\Gamma = 0.24 - 0.25$) in just these studies. This clearly demonstrates the usefulness of the ratio $\Gamma$ in characterizing a critical phase transition. In the following, we analyze our expansivity data in terms of $\Gamma$ and $B^+ / B^- (= A^+ / A^-)$, and then we make a direct comparison with ${}^4$He.

To check for the nearly logarithmic divergence expected in the critical region, $\tilde{\alpha}_{b-a}$ is plotted versus $log |T - T_c|$ (or $log|t|$) for both crystals in Fig. 2. Above $T_c$, $\tilde{\alpha}_{b-a}$ decreases linearly with $log |T - T_c|$ up to about $(T_c + 20 \, K)$, implying a very large critical region. Above this point the background contribution becomes important. Below $T_c$, nearly linear behavior is observed only for $0.2 \, K < |T - T_c| < 1 \, K$, and a marked downward curvature, not anticipated by Eq.s 1-2, is seen between $1 \, K < |T - T_c| < 10 \, K$. This behavior is clearly not due to the background, which causes the data to curve upward for $|T - T_c| > 20 \, K$. We attribute this downward curvature to the decrease of the jump component ($\Delta \tilde{\alpha}_{b-a}$) as $T$ decreases, and, as we show below, this behavior can be modeled very well with a conventional 'mean-field' type temperature dependence, i.e. $\Delta \tilde{\alpha}_{b-a}(T) = \Delta \tilde{\alpha}_{b-a}(t_o) [T / T_c]^n$. A similar effect is also seen in ${}^4$He at comparable reduced temperatures [21] and nearly all specific heat analyses of HTSCs also incorporate this type of term, see e.g. [1-3,5,6]. Here, we use this term to obtain a more accurate value for $B^-$ by extrapolating to small t values.

In order to determine $B^+$, $B^-$, $\Delta \tilde{\alpha}_{b-a}$ and, thus, $\Gamma$, the data were fit to a ln|t| term above $T_c$ and to a $ln |t|$ term plus the $\Delta \tilde{\alpha}_{b-a}(t_o) [T / T_c]^n$ term below $T_c$. $B^\pm$ are just the prefactors of the $ln |t|$ terms and $\Delta \tilde{\alpha}_{b-a}$ is the vertical distance between these terms at $t_o$. The

solid lines in Fig. 2a- b are the best fits with $T_c$, $B^+$, $B^-$, $\Delta\tilde{\alpha}_{b-a}$, $n$ and an arbitrary constant background as free parameters. The fitting range, indicated by the white lines in Fig. 2, was approximately $0.2\ K < |T - T_c| < 9\ K$. The dotted lines in Fig. 2 a-b represent the $B^-\ ln|t|$ terms. Since the data closest to $T_c$ are the most significant, a $1/|t|^2$ weighting was used. As can be seen in Fig. 2a-b, an excellent description of the expansivity data for both crystals is obtained using this logarithmic description, which clearly indicates that $\alpha \approx 0$. Both crystals show nearly identical behavior, and deviations from the fits are seen only for $|T - T_c| < 0.2\ K$, demonstrating the sharpness of the transitions. $T_c$ values are shown in Fig. 2. The value for $n$ was found to be close to 2 for both crystals. For comparison, the solid (dotted) lines in Fig. 2c represent the best fit using 3D-Gaussian fluctuations ($A^{\pm}/|t|^{-0.5}$) plus a 'mean-field' anomaly for crystal SRL-2. This type of description clearly does not fit our data close to $T_c$; the residuals are a factor of 50 larger than for the critical fits. Such a clear distinction between critical and Gaussian behavior is made possible here, because the additional degrees of freedom of a background are absent.

Fig. 3 shows the least-square residuals of the critical ($ln\ |t|$) fits as a function of the universal amplitude ratio $B^+ / B^-$. The residuals clearly exhibit a minimum at $B^+ / B^- \approx 1$ for both crystals, in good agreement with the value found for $^4$He (1.054) [23]. The $\Gamma$ values of our data, on the other hand, are significantly smaller ($\Gamma = 0.184 \pm 0.005$ and $0.187 \pm 0.005$, for crystals SRL-1 and SRL-2, respectively) than the value for $^4$He ($\Gamma = 0.24$-$0.25$), the significance of which will be discussed in more detail below. The scaling exponent $\alpha$ can be obtained directly through Eq. 3 with the values of $\Gamma$ and $A^+ / A^-$. Assuming that $A^+ / A^- = 1.0 \pm 0.1$ (see Fig. 3), constrains $\alpha$ to $\alpha = 0 \mp 0.018$. Alternatively, if we fix $\alpha$ to the $^4$He value ($\alpha = -0.01285$), we obtain $A^+ / A^- = 1.072 \pm 0.001$, which is also very close to the $^4$He value of $A^+ / A^- = 1.054$. Thus, we conclude that both $A^+ / A^-$ and $\alpha$ of $YBa_2Cu_3O_{7-\delta}$ are very close to the values expected in the 3D-XY model. We note that for a three-component order-parameter in three - dimensions (3D-Heisenberg), the values of both $A^+ / A^- \approx 1.3$ and $\alpha = -0.121$ are considerably different from our values.

We now briefly discuss the possible significance of the deviation of the present $\Gamma$ value from that of $^4$He. In Fig. 2 a-b, this deviation is graphically illustrated by plotting the scaling curves expected from the $\Gamma$ value of $^4$He for T < T$_c$. Here, we have forced the $B^+\ ln|t|$ term of $^4$He and of our data to be identical and have taken $B^+ / B^- = 1.054$ for $^4$He [23]. The discrepancy between $^4$He and YBa$_2$Cu$_3$O$_{7-\delta}$ is clearly seen. Universality means that $\alpha$, $A^+ / A^-$ and $\Gamma$ are constant for a given universality class. For $^4$He, universality has been convincingly demonstrated by showing that scaling is unaffected by both pressure [16] and $^3$He impurities [25]. In both these studies $\Gamma$ was always found to be around *0.23-0.25*. In contrast, for YBa$_2$Cu$_3$O$_{7-\delta}$ there is evidence from specific heat data of YBa$_2$Cu$_3$O$_{7-\delta}$ that the shape of the anomaly (i.e. $\Gamma$ value) is a strong function of the O-content even within the 90 K - plateau [5,24]. This is unexpected behavior for 3D-XY scaling, since, according to the Harris criterion [26], the scaling behavior in the 3D-XY class should be unaffected by small amounts of impurities (e.g. O-vacancies). Here it is interesting to note that both specific heat [6] and thermal expansion [8] data of the highly anisotropic Bi$_2$Sr$_2$CaCu$_2$O$_{8+x}$ compound find a nearly symmetric anomaly at $T_c$ with practically no jump component, i.e. $\Gamma$ is very large. It therefore appears to us that YBa$_2$Cu$_3$O$_{7-\delta}$, as well as other HTSCs, show deviations from pure 3D-XY scaling as observed in $^4$He ($\Gamma \approx 0.24$) in the experimentally accessible region. Possible reasons for this deviation may have to do with the structural anisotropy of HTSCs, a dimensional cross-over, or plane-chain coupling in YBa$_2$Cu$_3$O$_{7-\delta}$. We are currently investigating this effect in more detail by varying the O-content of our crystals.

In conclusion, the thermal expansivity of optimally doped YBa$_2$Cu$_3$O$_{7-\delta}$ single crystals provides unambiguous evidence for critical fluctuations over a surprisingly large temperature range of at least $\left| T - T_c \right| \leq 10 \ K$. Gaussian fluctuations are ruled out in this temperature range. The critical fit yields practically equal fluctuation amplitudes above and below the transition ($A^+ / A^- = 0.9 - 1.1$), which restrains the leading exponent to $/\alpha/ \leq 0.018$. These values are quite consistent with 3D-XY scaling as observed in $^4$He. The newly introduced universal quantity $\Gamma$ differs somewhat from that found for $^4$He, possibly suggesting a small deviation from the purely isotropic 3D-XY universality class.

**Acknowledgments**

On of the authors (C.M.) is grateful to Rolf Heid, Alain Junod, and Gerry Lander for useful discussions. This work was partially supported by NEDO, Japan.

**Figures**

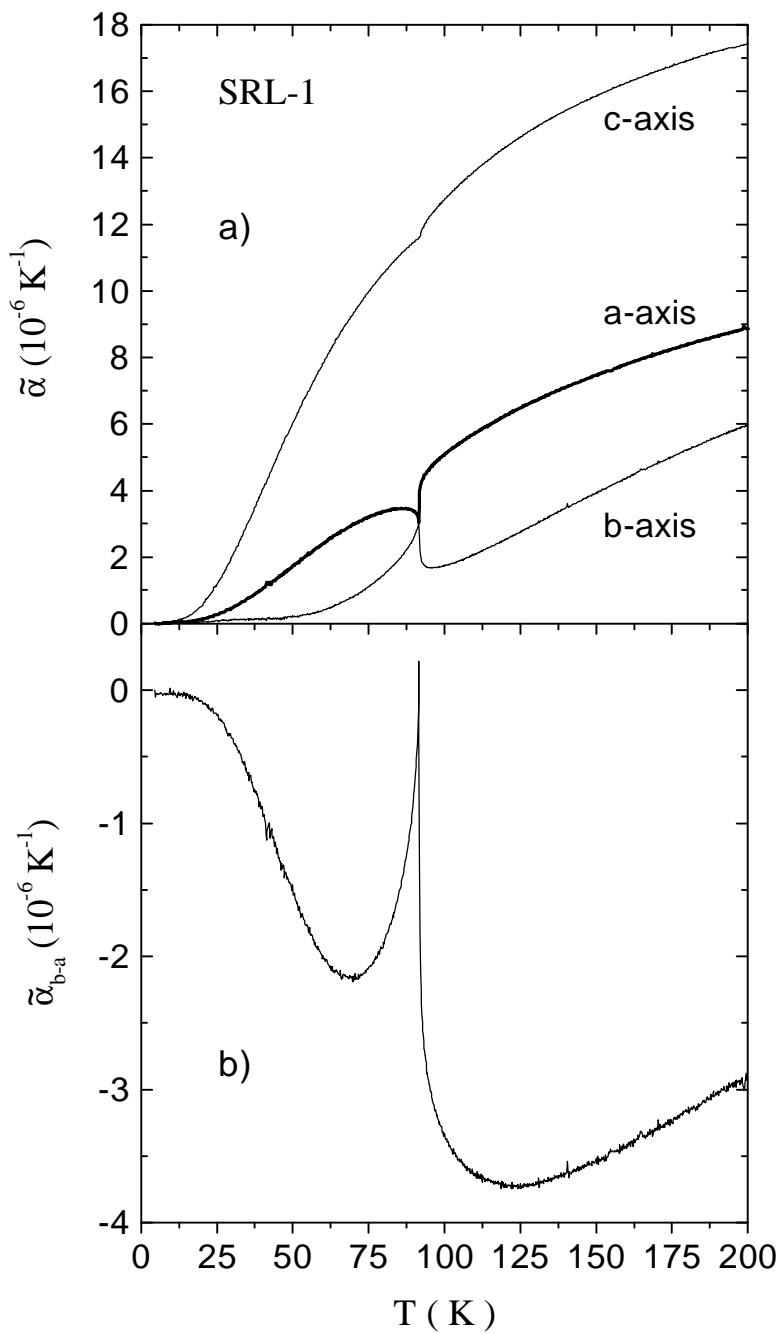

Fig. 1. a) Linear thermal expansivity $\tilde{\alpha}_i$ versus temperature for the three orthorhombic axes of crystal SRL-1. b) Difference between the expansivities along b- and a-axes ($\tilde{\alpha}_{b-a}$). $\tilde{\alpha}_{b-a}$ has no particular physical significance, but it is very useful for examining fluctuations because the anomalies in $\tilde{\alpha}_b$ and $\tilde{\alpha}_a$ are added, whereas the background is reduced. The anomaly in b) clearly has a $\lambda$-shape similar to the specific heat of $^4$He [21] and is about of equal magnitude as the slowly varying background.

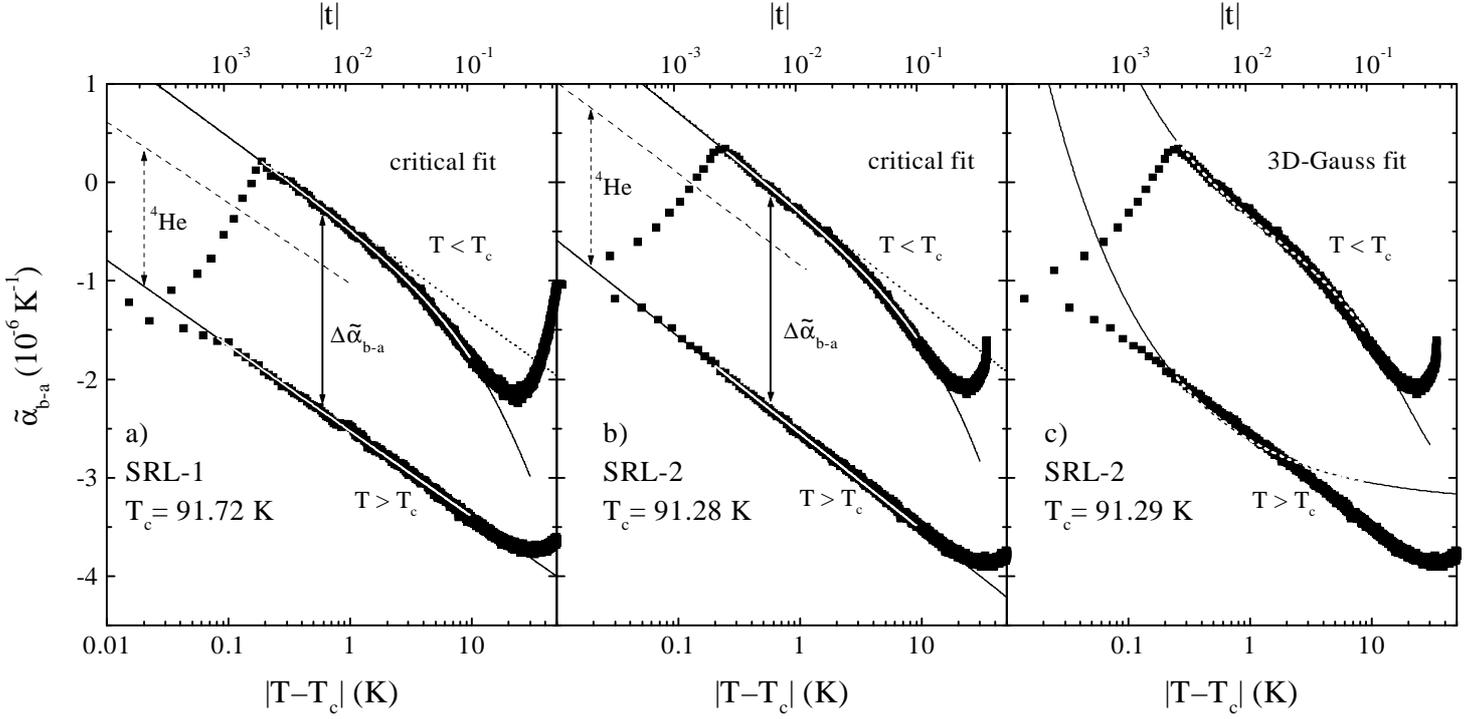

Fig. 2. $\tilde{\alpha}_{b-a}$ versus reduced temperature $t \equiv |T / T_c - 1|$ (or $|T - T_c|$) in a semi-log representation. Data (solid squares) of crystals a) SRL-1 and b) SRL-2. The black (white) lines in a) - b) represent the best fits to the critical model and c) shows the best 3D-Gauss fit for sample SRL-2. The white part of the lines in a) - c) represents the fitted region. Also shown in a) - b) are the jump components $\Delta \tilde{\alpha}_{b-a}$, the $B^- \ln|t|$ terms (dotted lines) and, for comparison, the scaling of $^4$He below $T_c$. (see text for details)

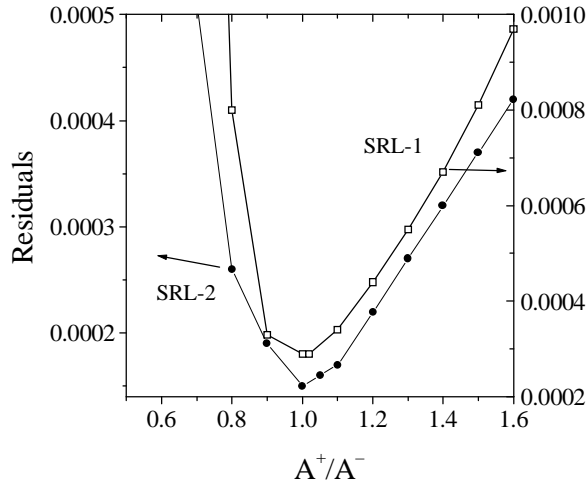

Fig. 3. Least-square residuals of the critical model fits as a function of the universal amplitude ratio $B^+ / B^-$ ($= A^+ / A^-$) for both crystals.